\documentclass[a4paper,11pt]{article}
\pdfoutput=1
\usepackage{jheppub}
\usepackage{amsmath,amssymb,latexsym}
\usepackage{braket}
\usepackage{bm}
\usepackage{booktabs}
\usepackage{setspace}
\usepackage{hepunits}
\usepackage[svgnames]{xcolor}

\usepackage{mciteplus}

\newcommand{\nn}{\nonumber}

\newcommand{\mG}{\mathcal{G}}
\newcommand{\aP}{\left\langle P \right\rangle}

\allowdisplaybreaks

\def\eref#1{(\ref{#1})}
\newcommand{\bea}{\begin{eqnarray}}
	\newcommand{\eea}{\end{eqnarray}}
\newcommand{\bean}{\begin{eqnarray*}}
	\newcommand{\eean}{\end{eqnarray*}}

\title{
Iteratively Reduce Auxiliary Scalar Product in Multi-loop Integrals
}

\author[a]{Jiaqi Chen}

\affiliation[a]{Beijing Computational Science Research Center, Beijing 100084, China}

\emailAdd{jiaqichen@csrc.ac.cn}

\abstract{In this paper, we construct a uniform formula that can iteratively reduce all auxiliary scalar product numerators of arbitrary multi-loop Feynman integrals. Integrals with such numerators commonly appear in Integration-By-Parts (IBP) relations. This formula is constructed with the generalized Sylvester's determinant identity. Compared to that using only traditional IBP reduction method, the combination of the formula and the traditional IBP method shows a significant speed-up.}

\begin{document}

\maketitle


\section{Introduction}

The integration-by-parts (IBP) reduction\cite{Chetyrkin:1981qh,Laporta:2000dsw} of Feynman integrals plays a central role in the computation of perturbative quantum field theory. It can be used to reduce a large number of integrals in the calculation of amplitudes into a linear combination of a few master integrals. The IBP reduction also generate the differential equation for evaluation of the master integrals. The precision improvement of current high-energy experiments requires more accurate theoretical predictions, so computations of high-loop or multi-scale progresses are needed. The computational resources required for reduction in these problems are also increasing. This has inspired people to develop many methods to improve the efficiency of reduction  \cite{Bern:1994zx,Bern:1994cg,Britto:2004nc,Britto:2005ha,Ossola:2006us,Gluza:2010ws,Peraro:2019svx,Chestnov:2022alh,Mastrolia:2018uzb,Frellesvig:2019kgj,Frellesvig:2019uqt,Weinzierl:2020xyy,Mizera:2019ose,Frellesvig:2020qot,Liu:2018dmc,Guan:2019bcx,Larsen:2015ped,Larsen:2016tdk,Zhang:2016kfo,Georgoudis:2016wff,Georgoudis:2017iza,Bohm:2017qme,Bohm:2018bdy,Bendle:2019csk,Boehm:2020zig,Bendle:2021ueg,Feng:2022uqp,Feng:2021enk,Hu:2021nia,Feng:2022rwj,Feng:2022rfz,Feng:2022iuc} (not only based on IBP method).

Recently, the strategy of iteration at the sector-level is proposed \cite{Chen:2022jux}. The strategy is to find these reduction relations that can iteratively reduce any integral in the top-sector to master integrals in the top-sector and keep the full information of the sub-sectors.
The matrix constructed by the IBP relations is sparse. Furthermore, it has repeated similar structure and redundant equations. Iterative reduction can take full advantage of these properties and avoid redundancy. Finding such iterative structures may also help investigate the mathematical structure of Feynman integrals and amplitude in the future. The current work is based on the syzygy and module intersection \cite{Gluza:2010ws,Schabinger:2011dz,Larsen:2015ped,Larsen:2016tdk,Zhang:2016kfo,Bohm:2017qme,Bohm:2018bdy,Bendle:2019csk,Boehm:2020zig} in Baikov representation \cite{Baikov:1996iu}. For one-loop cases \cite{Chen:2022jux}, people have suggested two methods to find such iterative reduction relations. One is to use computational algebraic geometry, which works case by case for each sector. This method is more likely to be successfully applied to a wider range of situations. Another method is to directly construct uniform formulas, which is more elegant and may keep more information.

In this paper, the multi-loop cases are investigated. While multi-loop cases are much more complicated than one-loop cases, we found a part of iterative reduction can also be achieved by a uniform formula. This part is reducing integrals with auxiliary scalar product (ASP) type numerators to integrals without ASP.

In this paper, the momenta linear independent of external momenta are denoted as auxiliary vectors. The scalar products of auxiliary vectors and loop momenta are denoted as ASP. 
Integrals with ASP-type numerators commonly arise in IBP relations. When we generate an IBP relation of a sector, propagators may be canceled in some terms, thus the ISP of this sector may becomes an ASP of the sub-sector in these terms. For example, let us consider generating an IBP in the top-sector of sunset:
\begin{align}
 \int d^dl_2  \int d^dl_1~ l_2^\mu\cdot  \frac{\partial}{\partial l_1^\mu }\left( \frac{l_2 \cdot p_1}{(l_1^2-m^2) (l_2^2-m^2) (l_1+l_2-p_1)^2}\right)=0.
\end{align}
Terms such as
\begin{align}
\int d^dl_2  \int d^dl_1  \frac{l_2 \cdot p_1}{(l_1^2-m^2)^2 (l_2^2-m^2) }
\end{align}
appear in this IBP relation. Since this is an integral in vacuum sector without external momentum, $p_1$ is an auxiliary vector and $l_2 \cdot p_1$ is an ASP.

Integrals with ASP-type numerators can also arise when people transform the tensor integrals into scalar integrals. People can achieve this by projecting the tensor integrals onto a complete basis of loop-momentum-independent tensor structures, e.g. as people have done in  \cite{Beenakker:2002nc}. People can also do this by reducing scalar integrals with auxiliary vectors and reading the tensor structure from the result \cite{Feng:2022uqp,Feng:2021enk,Hu:2021nia,Feng:2022rwj,Feng:2022rfz,Feng:2022iuc}. In both methods, integrals with ASP-type numerators can arise (some of which can also appear in the top-sector, unlike the IBP case).

In Sec.\ref{sec2}, we will give the notations of this paper, then we will construct the uniform formula and analysis its properties. This formula is constructed by a generalization of Sylvester's determinant identity whose proof is given in Appendix \ref{sec:app}. In Sec.\ref{sec:example}, we give an pedagogical example. In Sec.\ref{section-4}, we show the efficiency of a combination method in an example. In this combination method we reduce the ASP part iteratively and reduce the remaining part with widely used package FIRE6 \cite{Smirnov:2019qkx}. We also compare it to that using only the package. The combination method shows a significant speed-up.

\section{Notations and the uniform formula} \label{sec2}

Let's clarify the definition of ASP and the relevant notations first. For an L-loop Feynman integral, the loop momenta are denoted as $l_i$, and the E independent momenta of external legs are denoted as $p_i$. The integrand of Feynman integrals in this sector can be divided into three parts: propagators in the denominator (DP), irreducible scalar product (ISP) of loop momenta in the numerator such as $l_i \cdot l_j$ and $l_i \cdot p_j$, and auxiliary scalar product (ASP) such as $l_i \cdot R_j$ where $R_j$ is independent of $p_i$s and only appear in the numerator. They are denoted as
\begin{align}
	\text{DP: }&y_1=l_1^2-m_1^2,~~y_2=l_2^2-m_2^2, \cdots,y_N=(l_L-\cdots)^2-m_N^2 \nn\\
	\text{ISP: }&x_{1}=l_{i_1}\cdot p_{j_1} ,\cdots , x_M=l_{i_{M}}\cdot p_{j_{M}} . \nn\\
	\text{ASP: }&t_{1}=l_1\cdot R_1, \cdots,t_T= l_L
	\cdot R_v .
\end{align}
where $N+M=L(L+1)/2+LE$, N is the number of DP, and M is the number of ISP. $R_i$ are the v momenta which are independent of external momenta, and $T=L v$ is the number of ASP. The specific forms of ISP are determined by the propagators and have a degree of freedom to choose. DP, ISP, and ASP together are denoted as $z_i$ for short:
\begin{align}
	&z_1=y_1,\cdots,z_N=y_N,\nn\\
	&z_{N+1}=x_1,\cdots,z_{N+M}=x_M,\nn\\
	&z_{N+M+1}=t_1,\cdots,z_{N+M+T}=t_T.
\end{align}
The Feynman integrals are denoted as
\begin{align}
	\text{I}_{\{\bm{n}\};\{\bm{m}\};\{\bm{r}\}}&\equiv\text{I}_{n_1,\cdots,n_{N};m_{1},\cdots,m_{M};r_{1},\cdots,r_{T}} \equiv  \int\left[\prod_i^L \frac{d^d l_i}{i(\pi)^{d/2}}\right] \frac{\prod_{j=1}^{M} x_j^{m_j}\prod_{k=1}^{T} t_k^{r_k}}{\prod_{i=1}^{N} y_i^{n_i}}\nn\\
	&=\int\left[\prod_i^L \frac{d^d l_i}{i(\pi)^{d/2}} \right] \prod_{j=1}^{N+M+T} \frac{1}{z_j^{a_j}} \equiv \text{I}_{\{\bm{a}\}}.
\end{align}
By transforming the integral variables of the Feynman integral from loop momenta $\prod_{i} d^dl_i$ to propagators $\prod_{i} dz_i$, we get the standard form of Baikov representation \cite{Baikov:1996iu,Chen:2022lzr}.
\begin{align}
&\text{I}_{\{\bm{a}\}}= C_{E+v}^L(d) \int  \left[\prod_i^{N+M+T} \frac{dz_i}{ z_i^{a_i}} \right] \frac{\mG^{(d-L-E-T-1)/2}}{\mathcal{K}^{(d-E-T-1)/2}}.
\end{align}
The $C_{E}^L(d)$ and the Gram determinant $\mathcal{K}$ of $p_i$s and $R_i$s is independent to integral variables, so we ignore them in our discussion of IBP relations. $\mathcal{G}$ is
\begin{align}
\mathcal{G}(\bm{y};\bm{x};\bm{t})=G(\bm{l},\bm{p},\bm{R})= G(l_1,\cdots,l_L,p_1,\cdots,p_E,R_1,\cdots,R_v)
\end{align}
with the Gram determinant function $G$ defined as
\begin{align}
G(q_1,\ldots,q_n) \equiv \det (q_i \cdot q_j) = \left|
\begin{array}{ccc}
	q_1.q_1 & \cdots & q_1.q_n \\
	\vdots & \ddots    & \vdots \\
	q_n.q_1 & \cdots & q_n.q_n \\
\end{array}
\right|\nn\\
G\left(
\begin{array}{ccc}
	q_1, & \cdots, & q_n \\
	k_1, & \cdots, & k_n \\
\end{array}
\right)  \equiv \det (q_i \cdot k_j) = \left|
\begin{array}{ccc}
	q_1.k_1 & \cdots & q_1.k_n \\
	\vdots & \ddots    & \vdots \\
	q_n.k_1 & \cdots & q_n.k_n \\
\end{array} \right|
\end{align}
We can use a set of polynomials $\aP=\{P_1,\cdots,P_{N+M+T},P_0\}$ which satisfy
\begin{align}
&\sum_{i=1}^{N+M+T}  (P_i  \partial_{z_i} \mG )  + P_0 \mG=0   \nn\\
&P_i=z_i \bar{P}_i~~\text{for $i\leq N$}~~~~~ P_i=\bar{P}_i~~\text{for $i> N$}   \label{eq:syzygy}
\end{align}
 to generate IBP relations so that they do not involve dimension shift and higher power propagators (in the denominator) \cite{Larsen:2015ped,Bendle:2019csk}. The IBP relation given by 
  \begin{align}
& C \int \left[\prod_{i=1}^{N+M+T}dz_i\right] \sum_{j=1}^{N+M+T} \partial_{z_j}  \left\{P_j \left[ \prod_i^{N+M+T} \frac{1}{ z_i^{a_i}} \right] \mG^{(d-\gamma)/2} \right\}=0,  \nn\\
&\gamma= L+E+T+1 ,
 \end{align}
leads to
 \begin{align}
&	C \int \left\{D_{\left\langle P \right\rangle} \bm{\cdot} \frac{1}{\prod_{i=1}^{N+M+T} z_i^{a_i}}   \right\}  \mathcal{G}(\bm{z})^{(d-\gamma)/2} \prod_{i=1}^{N+M+T} dz_i=0,   \label{eq:ibp}
\end{align}
where
\begin{align}
& D_{\left\langle P \right\rangle} \bm{\cdot} Q \equiv  - \sum_{i=1}^{N+M+T} \left[ \partial_{z_i} \left( P_i \bm{\cdot} Q  \right) \right]  + \frac{d-\gamma}{2} P_0   \bm{\cdot} Q .
\end{align}

Let us consider the situation that there is only one R at the beginning. In this case, we have
\begin{align}
\mG = \left|
\begin{array}{ccc|c|c}
     &      &        &          &  l_1\cdot R  \\
     &  l_i \cdot l_j &          &l_i\cdot p_j   &\vdots   \\
     &      &         &          &l_L\cdot R   \\ \hline
      & l_i\cdot p_j  &         &  p_i\cdot p_j &   p_i\cdot R    \\ \hline   
l_1\cdot R  & \cdots  &  l_L\cdot R  & p_i\cdot R   &R^2  \\
\end{array} \right| .
\end{align}
The ASP only appear in the $l_i\cdot R$ part, while the DP and ISP only appear in the parts of  $l_i\cdot l_j$ and $l_i\cdot p_j$. This property leads to a solution to \eref{eq:syzygy}:
\begin{align}
&\sum_{i=1}^{N+M}0 \cdot \partial_{z_i} \mG  + \sum_{k=1}^{v}  G_{(j,k)} \cdot \partial_{t_k} \mG + 2 G_{(j,R)} \mG =0 
 \nn\\
&G_{(j,k)} \equiv G\left(
\begin{array}{ccc}
l_j, & R, & \bm{p} \\
l_k, & R, & \bm{p} \\
\end{array}\right)~~~~G_{(j,R)} \equiv G\left(
\begin{array}{ccc}
l_j, &  \bm{p} \\
R, &  \bm{p} \\
\end{array}\right)  \nn\\
&\bm{p}= p_1,p_2,\cdots,p_E      \label{eq:keyeq}
\end{align}
This is the key equation of this paper. By taking \eref{eq:keyeq} into \eref{eq:ibp}, we get the uniform formula that can iteratively reduce the power $r_j$ of ASP for each $j$, so it can reduce all ASP. This equation is a generalization of Sylvester's determinant identity. We give a proof of this equation in Appendix.\ref{sec:app}. Notice that when we transform the variables in $\mG$ from scalar product of loop momenta to $\bm{y}$, $\bm{x}$, and $\bm{t}$, the ASP $\bm{t}$ only apppear in one cloumn and row, but ISP are not the case. This is the reason why this formula can not be used for iterative reduction of ISP.

There are several important properties of this solution. 
Firstly, since only the $\partial_{y_i}\mG$ terms may increase the power of $y_i$ in the denominator, the IBP relations generated by this solution aviod this problem. 
Secondly, let us notice that only first column and first row of  $G_{(j,k)}$ involve loop momenta, thus the determinant is a quadratic polynomial of $\bm{z}$. Similarly, $G_{(j,R)}$ is linear in $\bm{z}$s. By analysing these observations, you will find iterative relation does not increase the total power of the numerator $\sum_j m_j + \sum_k r_k$. 
Furthermore, $t_j t_k$ only appears in $G_{j,k} $ with constant coefficient, so after $\partial_{t_k}$ acting on $\frac{t_j t_k}{\prod z_i^{a_i}} $ in \eref{eq:ibp}, the terms with highest total power of ASP only come from  $\frac{t_j}{\prod z_i^{a_i}} $. This ensures that any generated iterative relation does not involve two or more different terms with the same highest total power of ASP, such as $\text{I}_{\{\bm{n}\};\{\bm{m}\};4,3}$ and $\text{I}_{\{\bm{n}\};\{\bm{m}\};5,2}$. Thus it can directly decrease the total power of $\bm{t}$, instead of translating the power of one ASP to another ASP. $\mG$ is quadratic polynomial of $t_i$, thus when taking $\bm{r}$ equals $\bm{0}$ in \eref{eq:ibp}, people will generate a relation that can safely reduce $\text{I}_{\{\bm{n}\};\{\bm{m}\};0,0\cdots,0,r_j=1,0,\cdots,0}$ to integrals without ASP. Otherwise, it will relate to terms with ASP in the denominator. We will see these properties more explicitly in the example in the next section.

Obviously, when there are more $R_i$,  people can iteratively reduce $R_i$ terms one by one. When we have reduced all $l_i\cdot R_v$ numerators, $\mG$ can be reduced to $G(\bm{l},\bm{p},R_1,\cdots,R_{v-1})$ for the remaining integrals, because they do not involve $R_v$.

\begin{align}
	\mG = \left|
	\begin{array}{ccc|c|ccc}
		&      &        &          &  l_1\cdot R_1 &  \cdots   &l_1\cdot R_v  \\
		&  l_i \cdot l_j &          &l_i\cdot p_j   &\vdots &\ddots &\vdots   \\
		&      &         &          &l_L\cdot R_v &\cdots &l_L\cdot R_v  \\	\hline
		& l_i\cdot p_j  &         &  p_i\cdot p_j &  & p_i\cdot R_j &    \\	\hline   
		l_1\cdot R_1  & \cdots  &  l_L\cdot R_1  &   & & &  \\
		\vdots  & \ddots  &  \vdots  & p_i\cdot R_j   & &R_i\cdot R_j &  \\
		l_1\cdot R_v  & \cdots  &  l_L\cdot R_v  &    & & & \\
	\end{array} \right| .
\end{align}

\section{Pedagogical example}\label{sec:example}
Let us consider sunset propagators with one external R
\begin{align}
	&z_1=l_1^2-m_1^2,~~~~z_2=l_2^2-m_2^2,~~~~z_3=(l_1+l_2+p_1)^2-m_3^2,\nn\\
	&z_4=l_1\cdot p_1,~~~~z_5=l_2\cdot p_1,\nn\\
	&z_6=l_1\cdot R,~~~~z_7=l_2\cdot R.
\end{align}
Taking \eref{eq:keyeq} with $j=1$ and $\text{I}_{\bm{1};m_1,m_2;r_1-1,r_2}$ into \eref{eq:ibp}, we get
\begin{align} 
&\text{I}_{\bm{1};m_1,m_2;r_1,r_2} = \frac{Q}{p_1^2 \left(d+r_1+r_2-3\right)} \nn\\
&Q= \left. \left(d+2 r_1+r_2-4\right) R\cdot p_1  \text{I}_{\bm{1};m_1+1,m_2;r_1-1,r_2}  -\left(r_1-1\right) R^2 \text{I}_{\bm{1};m_1+2,m_2;r_1-2,r_2} \right.\nn\\
&\left. -m_1^2 \left(r_1-1\right) \left(\left(R\cdot p_1\right){}^2-p_1^2 R^2\right) \text{I}_{\bm{1};m_1,m_2;r_1-2,r_2} +r_2\left( \left(R\cdot p_1\right){}^2-p_1^2  R^2\right) \text{I}_{\bm{1};m_1,m_2+1;r_1-1,r_2-1}   \right.\nn\\
&\left. -\frac{1}{2} r_2 \left(m_1^2+p_1^2\right) \left(p_1^2 R^2  -\left(R\cdot p_1\right){}^2\right) \text{I}_{\bm{1};m_1,m_2;r_1-1,r_2-1}   -r_2 R^2 \text{I}_{\bm{1};m_1+1,m_2+1;r_1-1,r_2-1}  \right.\nn\\
&\left. +r_2 \left( \left(R\cdot p_1\right){}^2-p_1^2 R^2\right) \text{I}_{\bm{1};m_1+1,m_2;r_1-1,r_2-1}+r_2 R\cdot p_1 \text{I}_{\bm{1};m_1,m_2+1;r_1,r_2-1}  \right.\nn\\
&\left.+ s.s.t. \right.\nn\\
& s.s.t.=-\left(r_1-1\right) \left(\left(R\cdot p_1\right){}^2-p_1^2 R^2\right) \text{I}_{0,1,1;m_1,m_2;r_1-2,r_2} \nn\\
&+\frac{1}{2} r_2 \left(\left(R\cdot p_1\right){}^2-p_1^2 R^2\right) \text{I}_{0,1,1;m_1,m_2;r_1-1,r_2-1} \nn\\
&+\frac{1}{2} r_2 \left(\left(R\cdot p_1\right){}^2-p_1^2 R^2\right) \text{I}_{1,0,1;m_1,m_2;r_1-1,r_2-1}\nn\\
& -\frac{1}{2} r_2 \left(\left(R\cdot p_1\right){}^2-p_1^2 R^2\right) \text{I}_{1,1,0;m_1,m_2;r_1-1,r_2-1}
\end{align}
where s.s.t is short for sub-sector terms. As the analysis in the last section, the total power of numerators and the power of each denominator in each term do not increase either. It is a second-order iterative relation respect to $r_1+r_2$, but the $r_1-1$ and $r_2$ factor ensure the iterative reduction of the power $r_1$ can be safely accomplished when $r_1=1$ or $r_2=0$, and does not involve integrals with $r_1=-1$ or $r_2=-1$. After applying this equation, there are only integrals like $\text{I}_{\bm{1};m_1,m_2,0,r_2}$ and s.s.t remained. Then, we should take $j=2$ in \eref{eq:keyeq} and generate the iterative relation for $r_2$, which gives
\begin{align} 
&\text{I}_{\bm{1};m_1,m_2;0,r_2}=\frac{Q}{p_1^2 \left(d+r_2-3\right)}\nn\\
Q&=\left(d+2 r_2-4\right) R.p_1 \text{I}_{\bm{1};m_1,m_2+1,0,r_2-1}-m_1^2 \left(r_2-1\right) \left(\left(R.p_1\right){}^2-p_1^2 R^2\right) \text{I}_{\bm{1};m_1,m_2,0,r_2-2}\nn\\
&-\left(\left(r_2-1\right) R^2 \text{I}_{\bm{1};m_1,m_2+2;0,r_2-2}\right) + s.s.t\nn\\
&s.s.t=-\left(r_2-1\right) \left(\left(R.p_1\right){}^2-p_1^2 R^2\right) \text{I}_{1,0,1;m_1,m_2;0,r_2-2}.
\end{align}
This iterative relation finishes the reduction of ASP of the sunset sector.

\section{Comparison of efficiency and discussions}\label{section-4}
Let us consider a 2-loop 2-point sector with one auxilary vector R:
\begin{align}
	&z_1=l_1^2-m_1^2,~~~~z_4=(l_1-p_1)^2-m_2^2,~~~~z_3=l_2^2-m_3^2,~~~~z_4=(l_1+l_2)^2-m_4^2\nn\\
	&z_5=l_2\cdot p_1,\nn\\
	&z_6=l_1\cdot R,~~~~z_7=l_2\cdot R.
\end{align}
Integrals like $\text{I}_{1,1,1,1,0,-r_1,-r_2}$ commonly appears in many IBP relations of 2-loop 3-point sectors whose external momenta are $p_1$ and $R$.

In this section, we are going to compare the efficiency of reducing such integrals in two methods. The first method is to use the package FIRE6\cite{Smirnov:2019qkx}, while another method is to use the iterative formulas to reduce the two ASP numerators and then give it to FIRE6. In the FIRE6 (c++ version) part, the option "threads" for parallel computing is chosen to be 20. The iterative part is naively calculated in Mathematica without parallel computing. The CPU is Intel Xeon(R) Silver 4210R, $2.40\text{GHz}\times 40$.

For the rank-r reduction, we will give the set of all $I_{1,1,1,1,0,-r_1,-r_2}$ which satisfy $r_1+r_2=r$.
\begin{table}[ht]
	\centering
		\caption{Comparison}\label{tab:compare}
\begin{tabular}{ccccccc} %
	\toprule 
	r & 1  & 2 &3  &4 &5& 6 \\ 
	\midrule 
	FIRE6 only & 10(34) s  & 16(56) s  & 21(70) s & 27(78) s & 39(108)s & 75(164) s \\ 
	Iterative & $6*10^{-4}$ s  &  $3*10^{-3}$ s &  $1*10^{-2}$ s &  $4*10^{-2} $s & $0.3$ s & $1.6$ s \\ 
	FIRE6 & 4(15) s  & 5(17) s & 5(17) s & 6(20)s &6(22) s &9(27) s \\ 
	\bottomrule   
\end{tabular}
For the FIRE6 part, the time out of brackets are the total time, and the time in the brackets are the thread time, both given by FIRE6 itself.
\end{table}
Obviously, the combination of the traditional method and the iterative reduction of ASP is much quicker. I want to remind you that we should not just compare the total time of the two. The time of "FIRE6 only" minus the time of "FIRE6" can be roughly regarded as the time of reducing ASP in traditional method. Compare it to the time of "iterative", you will find the true improvement of iterative method.

Although the iterative part of the computation seems growing too  rapidly, which may worry people. We argue that this is because the algorithm is not optimized enough. If we combine it with a improved Laporta's algorithm to aviod repeated calculation of the low rank terms, and work in C++ \footnote{From the practice of FIRE6, it is well-known that the C++ version is much more effective than the Mathematica version}, it will have a better performance.

There is another advantage of iterative reduction. For higher-point, muti-loop, and multi-scale cases, the polynomial coefficients of master integrals will become much more complicated than shown in the examples. In iterative reduction, These polynomials can be kept always together. For example, people can denote them as some parameters in the process of reduction, which will make the expression much shorter and save the memory of the computer. When they need to be taken a value, people can also calculate the value of those parameters first, and then take it into the expression. While in the traditional reduction, you need to evaluate a messier and huger expression. 

\section{Summary and Outlook}\label{sec:so}
We construct a uniform formula that can iteratively reduce all ASP numerators and meanwhile does not increase the total power of ISP and ASP (in numerators) and the power of each DP. We compare the iterative reduction to traditional reduction, it shows a significant speed-up. 

With ASP part solved, we can focus on the reduction of ISP part in the future. This work highly suggests people to develop the iterative reduction of ISP part, which may improve the ability of physicists to perform high-precision calculations. Although an elegant uniform formula for iteratively reducing the ISP-type numerators may not exist, developing an algorithm to find iterative relations of ISP in the module intersection is possible. We will explore this in the future.

\section*{Acknowledgments}
This work is supported by  Chinese NSF funding under Grant No.11935013, No.11947301, No.12047502 (Peng Huanwu Center).

\appendix 
\section{Proof of the generalized Sylvester's determinant identity}\label{sec:app}
Let us consider a general matrix $X$ 
\begin{align}
X  =\left[ \setlength{\arraycolsep}{5pt}\begin{array}{c}
X_{i,j}  \\
\end{array} \right]  = \left[
\setlength{\arraycolsep}{16pt}\begin{array}{c|c}
\\
A  & B  \\
& \\  \hline
 & \\
C  &  K  \\
 & \\
\end{array} \right],
\end{align}
where A is a $m \times m$ matrix and $K$ is a $n \times n$ matrix. B and C can be canceled by $K$. Meanwhile, X will be transformed to $\widetilde{X}$ with its determinant unchanged.
\begin{align}
&\widetilde{X} = \left[
\setlength{\arraycolsep}{16pt}\begin{array}{c|c}
\\
\widetilde{A}  & 0  \\
 & \\ 
 \hline
 & \\
0  &  K  \\
 & \\
\end{array} \right]=\left[
\setlength{\arraycolsep}{8pt}\begin{array}{c|c}
1  & -B.K^{-1}  \\ 
\hline
0  &  1  \\
\end{array} \right].X.\left[
\setlength{\arraycolsep}{8pt}\begin{array}{c|c}
1  & 0  \\ 
\hline
-K^{-1}.C  & 1  \\
\end{array} \right],\nn\\
&\widetilde{A}=A-B.K^{-1}.C  ~ .
\end{align}
Obviously, $|X|=|\widetilde{X}|=|\widetilde{A}||K|$. Now, let us consider a Laplace expansion to $|\widetilde{A}|$.
\begin{align}
&\widetilde{X} = \left[
\setlength{\arraycolsep}{4pt}\begin{array}{c|ccc|ccc}
\widetilde{A}_{1,1}& & \cdots & & & &\\ 
\hline
\widetilde{A}_{2,1}& & & & & &\\
\vdots& ~& \widetilde{A}^{[1,1]}  &~ & & &\\
\widetilde{A}_{m,1}& & & & & &\\ 
\hline
& & & &  & &\\
& & & & ~ &K  &~   \\
& & & &  & &\\
\end{array} \right]\nn\\
|K| |\widetilde{X}|=&\sum_{j=1}^m \left[|K|~\widetilde{A}_{j,1}\right] \left[(-1)^{j+1}|\widetilde{A}^{[j,1]}| ~|K|\right]=(-1)^{j+1}|K^{(j,1)}| |\widetilde{X}^{[j,1]}|,
\end{align}
where the superscript $[i,j]$ represents the algebraic cofactor of $i,j$. Since
\begin{align}
(-1)^{j+1} |\widetilde{X}^{[j,1]}|=\frac{\partial}{\partial \widetilde{X}_{j,1}} |\widetilde{X}|=\frac{\partial}{\partial X_{j,1}}|\widetilde{X}|=\frac{\partial}{\partial X_{j,1}}|X|=(-1)^{j+1} |X^{[j,1]}|,
\end{align}
and similiarly, 
\begin{align}
|K^{(j,1)}|\equiv \widetilde{A}_{j,1}|K|=\left|
\setlength{\arraycolsep}{4pt}\begin{array}{c|ccc}
A_{j,1}& B_{j,m+1}&\cdots&B_{j,m+n}\\ 
\hline
C_{m+1,1}&  & &\\
\vdots& ~ &K  &~   \\
C_{m+n,1}&  & &\\
\end{array} \right|,
\end{align}
we have
\begin{align}
\sum_{j=1}^m |K^{(j,1)}| \frac{\partial}{\partial X_{j,1}}|X| = |K||X|.
\end{align}
With a similiar proof, it can be generalized to
\begin{align}
\sum_{k=1}^m |K^{(k,i)}| \frac{\partial}{\partial X_{k,j}}|X| = \delta_{i,j}|K||X|.
\end{align}
While $|X|$ is the determinant of a symmetric matrix, it becomes
\begin{align}
\sum_{k=1}^m |K^{(k,i)}| \frac{\partial}{\partial X_{k,j}}|X| = 2 \delta_{i,j}|K||X|. \label{eq:sysy}
\end{align}
For
\begin{align}
&|X|=\mG=G(\bm{l},\bm{p},\bm{R})=(-1)^{j-1}G\left(
\begin{array}{cc|cc}
&\bm{l}, &R, & \bm{p} \\
R & \bm{l}_{\hat{j}},& l_j & \bm{p} \\
\end{array}
\right) \nn\\
&\bm{l}_{\hat{j}} \equiv l_1,\cdots,l_{j-1},l_{j+1},\cdots,l_L,
\end{align}
we have
\begin{align}
&X_{k,1}=l_k.R~~~~K=G\left(
\begin{array}{cc}
R, & \bm{p} \\
l_j & \bm{p} \\
\end{array}
\right)=G_{(j,R)}  ~~~~  K^{(k,1)}=G\left(
\begin{array}{ccc}
l_k,R, & \bm{p} \\
R,l_j, & \bm{p} \\
\end{array}
\right)=-G_{(j,k)}~.
\end{align}
This is the equation in \eref{eq:keyeq}.

\bibliographystyle{JHEP}
\bibliography{references}

\providecommand{\href}[2]{#2}\begingroup\raggedright\begin{thebibliography}{10}

\bibitem{Chetyrkin:1981qh}
K.G.~Chetyrkin and F.V.~Tkachov, \emph{{Integration by Parts: The Algorithm to
  Calculate beta Functions in 4 Loops}},
  \href{https://doi.org/10.1016/0550-3213(81)90199-1}{\emph{Nucl. Phys. B}
  {\bfseries 192} (1981) 159}.

\bibitem{Laporta:2000dsw}
S.~Laporta, \emph{{High precision calculation of multiloop Feynman integrals by
  difference equations}},
  \href{https://doi.org/10.1142/S0217751X00002159}{\emph{Int. J. Mod. Phys. A}
  {\bfseries 15} (2000) 5087}
  [\href{https://arxiv.org/abs/hep-ph/0102033}{{\ttfamily hep-ph/0102033}}].

\bibitem{Bern:1994zx}
Z.~Bern, L.J.~Dixon, D.C.~Dunbar and D.A.~Kosower, \emph{{One loop n point
  gauge theory amplitudes, unitarity and collinear limits}},
  \href{https://doi.org/10.1016/0550-3213(94)90179-1}{\emph{Nucl. Phys. B}
  {\bfseries 425} (1994) 217}
  [\href{https://arxiv.org/abs/hep-ph/9403226}{{\ttfamily hep-ph/9403226}}].

\bibitem{Bern:1994cg}
Z.~Bern, L.J.~Dixon, D.C.~Dunbar and D.A.~Kosower, \emph{{Fusing gauge theory
  tree amplitudes into loop amplitudes}},
  \href{https://doi.org/10.1016/0550-3213(94)00488-Z}{\emph{Nucl. Phys. B}
  {\bfseries 435} (1995) 59}
  [\href{https://arxiv.org/abs/hep-ph/9409265}{{\ttfamily hep-ph/9409265}}].

\bibitem{Britto:2004nc}
R.~Britto, F.~Cachazo and B.~Feng, \emph{{Generalized unitarity and one-loop
  amplitudes in N=4 super-Yang-Mills}},
  \href{https://doi.org/10.1016/j.nuclphysb.2005.07.014}{\emph{Nucl. Phys. B}
  {\bfseries 725} (2005) 275}
  [\href{https://arxiv.org/abs/hep-th/0412103}{{\ttfamily hep-th/0412103}}].

\bibitem{Britto:2005ha}
R.~Britto, E.~Buchbinder, F.~Cachazo and B.~Feng, \emph{{One-loop amplitudes of
  gluons in SQCD}},
  \href{https://doi.org/10.1103/PhysRevD.72.065012}{\emph{Phys. Rev. D}
  {\bfseries 72} (2005) 065012}
  [\href{https://arxiv.org/abs/hep-ph/0503132}{{\ttfamily hep-ph/0503132}}].

\bibitem{Ossola:2006us}
G.~Ossola, C.G.~Papadopoulos and R.~Pittau, \emph{{Reducing full one-loop
  amplitudes to scalar integrals at the integrand level}},
  \href{https://doi.org/10.1016/j.nuclphysb.2006.11.012}{\emph{Nucl. Phys. B}
  {\bfseries 763} (2007) 147}
  [\href{https://arxiv.org/abs/hep-ph/0609007}{{\ttfamily hep-ph/0609007}}].

\bibitem{Gluza:2010ws}
J.~Gluza, K.~Kajda and D.A.~Kosower, \emph{{Towards a Basis for Planar Two-Loop
  Integrals}}, \href{https://doi.org/10.1103/PhysRevD.83.045012}{\emph{Phys.
  Rev. D} {\bfseries 83} (2011) 045012}
  [\href{https://arxiv.org/abs/1009.0472}{{\ttfamily 1009.0472}}].

\bibitem{Peraro:2019svx}
T.~Peraro, \emph{{FiniteFlow: multivariate functional reconstruction using
  finite fields and dataflow graphs}},
  \href{https://doi.org/10.1007/JHEP07(2019)031}{\emph{JHEP} {\bfseries 07}
  (2019) 031} [\href{https://arxiv.org/abs/1905.08019}{{\ttfamily
  1905.08019}}].

\bibitem{Chestnov:2022alh}
V.~Chestnov, F.~Gasparotto, M.K.~Mandal, P.~Mastrolia, S.J.~Matsubara-Heo,
  H.J.~Munch et~al., \emph{{Macaulay Matrix for Feynman Integrals: Linear
  Relations and Intersection Numbers}},
  \href{https://arxiv.org/abs/2204.12983}{{\ttfamily 2204.12983}}.

\bibitem{Mastrolia:2018uzb}
P.~Mastrolia and S.~Mizera, \emph{{Feynman Integrals and Intersection Theory}},
  \href{https://doi.org/10.1007/JHEP02(2019)139}{\emph{JHEP} {\bfseries 02}
  (2019) 139} [\href{https://arxiv.org/abs/1810.03818}{{\ttfamily
  1810.03818}}].

\bibitem{Frellesvig:2019kgj}
H.~Frellesvig, F.~Gasparotto, S.~Laporta, M.K.~Mandal, P.~Mastrolia,
  L.~Mattiazzi et~al., \emph{{Decomposition of Feynman Integrals on the Maximal
  Cut by Intersection Numbers}},
  \href{https://doi.org/10.1007/JHEP05(2019)153}{\emph{JHEP} {\bfseries 05}
  (2019) 153} [\href{https://arxiv.org/abs/1901.11510}{{\ttfamily
  1901.11510}}].

\bibitem{Frellesvig:2019uqt}
H.~Frellesvig, F.~Gasparotto, M.K.~Mandal, P.~Mastrolia, L.~Mattiazzi and
  S.~Mizera, \emph{{Vector Space of Feynman Integrals and Multivariate
  Intersection Numbers}},
  \href{https://doi.org/10.1103/PhysRevLett.123.201602}{\emph{Phys. Rev. Lett.}
  {\bfseries 123} (2019) 201602}
  [\href{https://arxiv.org/abs/1907.02000}{{\ttfamily 1907.02000}}].

\bibitem{Weinzierl:2020xyy}
S.~Weinzierl, \emph{{On the computation of intersection numbers for twisted
  cocycles}}, \href{https://doi.org/10.1063/5.0054292}{\emph{J. Math. Phys.}
  {\bfseries 62} (2021) 072301}
  [\href{https://arxiv.org/abs/2002.01930}{{\ttfamily 2002.01930}}].

\bibitem{Mizera:2019ose}
S.~Mizera, \emph{{Status of Intersection Theory and Feynman Integrals}},
  \href{https://doi.org/10.22323/1.383.0016}{\emph{PoS} {\bfseries MA2019}
  (2019) 016} [\href{https://arxiv.org/abs/2002.10476}{{\ttfamily
  2002.10476}}].

\bibitem{Frellesvig:2020qot}
H.~Frellesvig, F.~Gasparotto, S.~Laporta, M.K.~Mandal, P.~Mastrolia,
  L.~Mattiazzi et~al., \emph{{Decomposition of Feynman Integrals by
  Multivariate Intersection Numbers}},
  \href{https://doi.org/10.1007/JHEP03(2021)027}{\emph{JHEP} {\bfseries 03}
  (2021) 027} [\href{https://arxiv.org/abs/2008.04823}{{\ttfamily
  2008.04823}}].

\bibitem{Liu:2018dmc}
X.~Liu and Y.-Q.~Ma, \emph{{Determining arbitrary Feynman integrals by vacuum
  integrals}}, \href{https://doi.org/10.1103/PhysRevD.99.071501}{\emph{Phys.
  Rev. D} {\bfseries 99} (2019) 071501}
  [\href{https://arxiv.org/abs/1801.10523}{{\ttfamily 1801.10523}}].

\bibitem{Guan:2019bcx}
X.~Guan, X.~Liu and Y.-Q.~Ma, \emph{{Complete reduction of integrals in
  two-loop five-light-parton scattering amplitudes}},
  \href{https://doi.org/10.1088/1674-1137/44/9/093106}{\emph{Chin. Phys. C}
  {\bfseries 44} (2020) 093106}
  [\href{https://arxiv.org/abs/1912.09294}{{\ttfamily 1912.09294}}].

\bibitem{Larsen:2015ped}
K.J.~Larsen and Y.~Zhang, \emph{{Integration-by-parts reductions from unitarity
  cuts and algebraic geometry}},
  \href{https://doi.org/10.1103/PhysRevD.93.041701}{\emph{Phys. Rev. D}
  {\bfseries 93} (2016) 041701}
  [\href{https://arxiv.org/abs/1511.01071}{{\ttfamily 1511.01071}}].

\bibitem{Larsen:2016tdk}
K.J.~Larsen and Y.~Zhang, \emph{{Integration-by-parts reductions from the
  viewpoint of computational algebraic geometry}},
  \href{https://doi.org/10.22323/1.260.0029}{\emph{PoS} {\bfseries LL2016}
  (2016) 029} [\href{https://arxiv.org/abs/1606.09447}{{\ttfamily
  1606.09447}}].

\bibitem{Zhang:2016kfo}
Y.~Zhang, \emph{{Lecture Notes on Multi-loop Integral Reduction and Applied
  Algebraic Geometry}},  12, 2016
  [\href{https://arxiv.org/abs/1612.02249}{{\ttfamily 1612.02249}}].

\bibitem{Georgoudis:2016wff}
A.~Georgoudis, K.J.~Larsen and Y.~Zhang, \emph{{Azurite: An algebraic geometry
  based package for finding bases of loop integrals}},
  \href{https://doi.org/10.1016/j.cpc.2017.08.013}{\emph{Comput. Phys. Commun.}
  {\bfseries 221} (2017) 203}
  [\href{https://arxiv.org/abs/1612.04252}{{\ttfamily 1612.04252}}].

\bibitem{Georgoudis:2017iza}
A.~Georgoudis, K.J.~Larsen and Y.~Zhang, \emph{{Cristal and Azurite: new tools
  for integration-by-parts reductions}},
  \href{https://doi.org/10.22323/1.290.0020}{\emph{PoS} {\bfseries RADCOR2017}
  (2017) 020} [\href{https://arxiv.org/abs/1712.07510}{{\ttfamily
  1712.07510}}].

\bibitem{Bohm:2017qme}
J.~B\"ohm, A.~Georgoudis, K.J.~Larsen, M.~Schulze and Y.~Zhang, \emph{{Complete
  sets of logarithmic vector fields for integration-by-parts identities of
  Feynman integrals}},
  \href{https://doi.org/10.1103/PhysRevD.98.025023}{\emph{Phys. Rev. D}
  {\bfseries 98} (2018) 025023}
  [\href{https://arxiv.org/abs/1712.09737}{{\ttfamily 1712.09737}}].

\bibitem{Bohm:2018bdy}
J.~B\"ohm, A.~Georgoudis, K.J.~Larsen, H.~Sch\"onemann and Y.~Zhang,
  \emph{{Complete integration-by-parts reductions of the non-planar hexagon-box
  via module intersections}},
  \href{https://doi.org/10.1007/JHEP09(2018)024}{\emph{JHEP} {\bfseries 09}
  (2018) 024} [\href{https://arxiv.org/abs/1805.01873}{{\ttfamily
  1805.01873}}].

\bibitem{Bendle:2019csk}
D.~Bendle, J.~B\"ohm, W.~Decker, A.~Georgoudis, F.-J.~Pfreundt, M.~Rahn et~al.,
  \emph{{Integration-by-parts reductions of Feynman integrals using Singular
  and GPI-Space}}, \href{https://doi.org/10.1007/JHEP02(2020)079}{\emph{JHEP}
  {\bfseries 02} (2020) 079}
  [\href{https://arxiv.org/abs/1908.04301}{{\ttfamily 1908.04301}}].

\bibitem{Boehm:2020zig}
J.~Boehm, D.~Bendle, W.~Decker, A.~Georgoudis, F.-J.~Pfreundt, M.~Rahn et~al.,
  \emph{{Module Intersection for the Integration-by-Parts Reduction of
  Multi-Loop Feynman Integrals}},
  \href{https://doi.org/10.22323/1.383.0004}{\emph{PoS} {\bfseries MA2019}
  (2022) 004} [\href{https://arxiv.org/abs/2010.06895}{{\ttfamily
  2010.06895}}].

\bibitem{Bendle:2021ueg}
D.~Bendle, J.~Boehm, M.~Heymann, R.~Ma, M.~Rahn, L.~Ristau et~al.,
  \emph{{Two-loop five-point integration-by-parts relations in a usable form}},
   \href{https://arxiv.org/abs/2104.06866}{{\ttfamily 2104.06866}}.

\bibitem{Feng:2022uqp}
B.~Feng, T.~Li, H.~Wang and Y.~Zhang, \emph{{Reduction of general one-loop
  integrals using auxiliary vector}},
  \href{https://doi.org/10.1007/JHEP05(2022)065}{\emph{JHEP} {\bfseries 05}
  (2022) 065} [\href{https://arxiv.org/abs/2203.14449}{{\ttfamily
  2203.14449}}].

\bibitem{Feng:2021enk}
B.~Feng, T.~Li and X.~Li, \emph{{Analytic tadpole coefficients of one-loop
  integrals}}, \href{https://doi.org/10.1007/JHEP09(2021)081}{\emph{JHEP}
  {\bfseries 09} (2021) 081}
  [\href{https://arxiv.org/abs/2107.03744}{{\ttfamily 2107.03744}}].

\bibitem{Hu:2021nia}
C.~Hu, T.~Li and X.~Li, \emph{{One-loop Feynman integral reduction by
  differential operators}},
  \href{https://doi.org/10.1103/PhysRevD.104.116014}{\emph{Phys. Rev. D}
  {\bfseries 104} (2021) 116014}
  [\href{https://arxiv.org/abs/2108.00772}{{\ttfamily 2108.00772}}].

\bibitem{Feng:2022rwj}
B.~Feng, J.~Gong and T.~Li, \emph{{Universal Treatment of Reduction for
  One-Loop Integrals in Projective Space}},
  \href{https://arxiv.org/abs/2204.03190}{{\ttfamily 2204.03190}}.

\bibitem{Feng:2022rfz}
B.~Feng, C.~Hu, T.~Li and Y.~Song, \emph{{Reduction with Degenerate Gram matrix
  for One-loop Integrals}},  \href{https://arxiv.org/abs/2205.03000}{{\ttfamily
  2205.03000}}.

\bibitem{Feng:2022iuc}
B.~Feng and T.~Li, \emph{{PV-Reduction of Sunset Topology with Auxiliary
  Vector}},  \href{https://arxiv.org/abs/2203.16881}{{\ttfamily 2203.16881}}.

\bibitem{Chen:2022jux}
J.~Chen and B.~Feng, \emph{{Module Intersection and Uniform Formula for
  Iterative Reduction of One-loop Integrals}},
  \href{https://arxiv.org/abs/2207.03767}{{\ttfamily 2207.03767}}.

\bibitem{Schabinger:2011dz}
R.M.~Schabinger, \emph{{A New Algorithm For The Generation Of
  Unitarity-Compatible Integration By Parts Relations}},
  \href{https://doi.org/10.1007/JHEP01(2012)077}{\emph{JHEP} {\bfseries 01}
  (2012) 077} [\href{https://arxiv.org/abs/1111.4220}{{\ttfamily 1111.4220}}].

\bibitem{Baikov:1996iu}
P.A.~Baikov, \emph{{Explicit solutions of the multiloop integral recurrence
  relations and its application}},
  \href{https://doi.org/10.1016/S0168-9002(97)00126-5}{\emph{Nucl. Instrum.
  Meth. A} {\bfseries 389} (1997) 347}
  [\href{https://arxiv.org/abs/hep-ph/9611449}{{\ttfamily hep-ph/9611449}}].

\bibitem{Beenakker:2002nc}
W.~Beenakker, S.~Dittmaier, M.~Kramer, B.~Plumper, M.~Spira and P.M.~Zerwas,
  \emph{{NLO QCD corrections to t anti-t H production in hadron collisions}},
  \href{https://doi.org/10.1016/S0550-3213(03)00044-0}{\emph{Nucl. Phys. B}
  {\bfseries 653} (2003) 151}
  [\href{https://arxiv.org/abs/hep-ph/0211352}{{\ttfamily hep-ph/0211352}}].

\bibitem{Smirnov:2019qkx}
A.V.~Smirnov and F.S.~Chuharev, \emph{{FIRE6: Feynman Integral REduction with
  Modular Arithmetic}},
  \href{https://doi.org/10.1016/j.cpc.2019.106877}{\emph{Comput. Phys. Commun.}
  {\bfseries 247} (2020) 106877}
  [\href{https://arxiv.org/abs/1901.07808}{{\ttfamily 1901.07808}}].

\bibitem{Chen:2022lzr}
J.~Chen, X.~Jiang, C.~Ma, X.~Xu and L.L.~Yang, \emph{{Baikov representations,
  intersection theory, and canonical Feynman integrals}},
  \href{https://doi.org/10.1007/JHEP07(2022)066}{\emph{JHEP} {\bfseries 07}
  (2022) 066} [\href{https://arxiv.org/abs/2202.08127}{{\ttfamily
  2202.08127}}].

\end{thebibliography}\endgroup

\end{document}